\documentclass[runningheads,english]{llncs}

\usepackage[mathlines]{lineno}
\usepackage[T1]{fontenc}
\usepackage[english]{babel}
\usepackage[utf8]{inputenc}
\usepackage{paralist}

\usepackage{amsmath}
\usepackage{amssymb}
\usepackage{xspace}
\usepackage{marcel_notation}
\usepackage[backgroundcolor=white,linecolor=black,textsize=scriptsize]{todonotes}

\usepackage{subfig}
\usepackage{multirow}
\usepackage[noabbrev]{cleveref}
\usepackage{wrapfig}

\usepackage[linesnumbered,ruled, vlined]{algorithm2e}
\SetKwInOut{Input}{Input}
\SetKwInOut{Output}{Output}

\newcommand{\north}{\textsc{North}\xspace}
\newcommand{\rome}{\textsc{Rome}\xspace}
\newcommand{\community}{\textsc{Community}\xspace}
\newcommand{\Oneplanar}{\textsc{1-Planar}\xspace}

\newcommand{\TriX}{\textsc{Triangulation+X}\xspace}

\newcommand{\Sloppy}{\textsc{Sloppy}\xspace}
\newcommand{\Medium}{\textsc{Medium}\xspace}
\newcommand{\Precise}{\textsc{Precise}\xspace}

\newcommand{\randomSampling}{\textsc{Random Sampling}\xspace}

\newcommand\frcos{\textsc{Fr+Cos}\xspace}
\newcommand{\rnd}{\textsc{Random}\xspace}
\newcommand{\stress}{\textsc{Stress}\xspace}
\newcommand{\crmin}{\textsc{Cr-small}\xspace}

\newcommand{\addgood}{$^\star$}
\newcommand\frcosRS{{\frcos}\addgood\xspace}
\newcommand{\rndRS}{{\rnd}\addgood\xspace}
\newcommand{\stressRS}{{\stress}\addgood\xspace}
\newcommand{\stressRSS}{{\stress}\addgood\addgood\xspace}
\newcommand{\crminRS}{{\crmin}\addgood\xspace}

 \newcommand{\Mod}[1]{\ (\mathrm{mod}\ #1)}

\newif\iflong
\longtrue

\DeclareMathOperator{\slope}{slope}

\title{A Greedy Heuristic for Crossing-Angle Maximization\thanks{Work
was partially supported by grant WA 654/21-1 of the German Research Foundation
(DFG).}
}

\author{Almut Demel 
	\and Dominik Dürrschnabel 
	\and Tamara Mchedlidze 
	\and Marcel Radermacher 
	\and Lasse Wulf}

\authorrunning{\xspace}

\institute{Department of Computer Science, Karlsruhe  Institute of Technology,
	Germany	\email{mched@iti.uka.de}, \email{radermacher@kit.edu}
}
\date{}

\begin{document}
\maketitle

\begin{abstract}
	The crossing angle of a straight-line drawing $\Gamma$ of a graph $G=(V, E)$
	is the smallest angle between two crossing edges in $\Gamma$. Deciding whether
	a graph $G$ has a straight-line drawing with a crossing angle of $90^\circ$ is
	$\cNP$-hard~\cite{JGAA-274}. We propose a simple heuristic to compute a
	drawing with a large crossing angle. The heuristic greedily selects the best
	position for a single vertex in a random set of points. The algorithm is
	accompanied by a speed-up technique to compute the crossing angle of a
	straight-line drawing. We show the effectiveness  of the heuristic in an
	extensive empirical evaluation.  Our heuristic was clearly the winning
	algorithm (CoffeeVM) in the Graph Drawing Challenge
	2017~\cite{10.1007/978-3-319-73915-1_44}.
\end{abstract}

\section{Introduction} \label{sec:intro} 

The \emph{crossing angle} $\Crangle(\Gamma)$ of a straight-line drawing $\Gamma$
is defined to be the minimum over all angles created by two crossing edges in
$\Gamma$.  The 24th edition of the annual \emph{Graph Drawing Challenge}, held
during the Graph Drawing Symposium, posed the following problem: Given a graph
$G$, compute a straight-line drawing $\Gamma$ on an integer grid that has
a large crossing angle. In this paper we present a greedy heuristic that starts
with a carefully chosen initial drawing and repeatedly moves a vertex $v$ to a
random point $p$ if this increases the crossing angle of~$\Gamma$. This
heuristic was the winning algorithm of the GD Challenge
'17~\cite{10.1007/978-3-319-73915-1_44}.

\subsubsection{Related Works}

A drawing of a graph is called \emph{RAC} if its minimum crossing angle is
$90^\circ$.  Deciding whether a graph has a straight-line RAC drawing is an
$\cNP$-hard problem~\cite{JGAA-274}.  Giacomo et al.\cite{DIGIACOMO2012132}
proved that every straight-line drawing of a complete graph with at least 12
vertices has a crossing angle of $\Theta(\pi / n)$.  Didimo et
al.~\cite{DBLP:journals/tcs/DidimoEL11} have shown that every $n$-vertex graph
that admits a straight-line RAC drawing has at most $4n-10$ edges. This bound is
tight, since there is an infinite family of graphs with $4n-10$ edges that have
straight-line RAC drawings.  Moreover they proved that every graph has a RAC
drawing with three bends per edge.  Arikishu et al.~\cite{ARIKUSHI2012169}
showed that any $n$-vertex graph that admits a RAC drawing with one bend or two
bends per edge has at most $6.5n$ and $74.2n$ edges, respectively.  For an
overview over further results on RAC drawings we refer to \cite{Didimo2013}.
Dujmović et al.\cite{Dujmovic:2010:NLA:1862317.1862320} introduced the concept
of $\alpha$AC graphs. A graph is \emph{$\alpha$AC} if it admits a drawing with
crossing angle of at least $\alpha$. For $\alpha > \pi/3$, $\alpha$AC graphs are
\emph{quasiplanar} graphs, i.e., graphs that admit a drawing without three
mutually crossing edges, and thus have at most $6.5n - 20$ edges.  Moreover,
every $n$-vertex $\alpha$AC graph with $\alpha \in (0, \pi /2)$ has at most
$(\pi / \alpha) (3n - 6)$ edges. 
Besides the theoretical work on this topic, there are a few force-directed
approaches that optimize the crossing angle in drawings of arbitrary
graphs~\cite{5635222,DBLP:journals/cj/ArgyriouBS13}, see Sec.~\ref{sec:forces}.  

\subsubsection{Contribution}
\label{sec:contribution}

We introduce a heuristic to increase the crossing angle in a given straight-line
drawing $\Gamma$ (Sec.~\ref{sec:random_sampling}). The heuristic is accompanied
by a speed-up technique to compute the pair of crossing edges in $\Gamma$ that
create the smallest crossing angle. In Sec.~\ref{sec:evalution} we give an
extensive evaluation of our heuristic. The evaluation is driven by three main research
questions:
\begin{inparaenum}[i)]
	\item What is a good parametrization of our heuristic?
	\item Does our heuristic improve the crossing angle of a given initial drawing?
	\item What is a good choice for an initial drawing?
\end{inparaenum}

\section{Preliminaries}

Let $\Gamma$ be a straight-line drawing of a graph $G=(V,E)$. Denote by $n$ and
$m$ the number of vertices and edges of $G$, respectively.
Let $e$ and $e'$
be two distinct edges of $G$. If $e$ and $e'$ have an interior intersection in
$\Gamma$, the function $\Crangle(\Gamma, e, e')$ denotes the smallest angle
formed by $e$ and $e'$ in $\Gamma$. In case that $e$ and $e'$ do not intersect,
we define $\Crangle(\Gamma, e, e')$ to be $\pi/2$. The \emph{local crossing
angle of a vertex $v$} is defined as the minimum angle of the edges incident to
$v$, i.e., ${\Crangle}(\Gamma, v) = \min_{e, uv \in E, e \not= uv}
\Crangle(\Gamma, e, uv)$. The \emph{crossing angle} of a drawing $\Gamma$ is
defined as $\Crangle(\Gamma) = \min_{e, e' \in E, e\not=e' } \Crangle(\Gamma, e,
e')$.  Let $\Delta x$ and $\Delta y$ be the difference of the x-coordinates and
the y-coordinates of the endpoints of $e$ in a drawing $\Gamma$.  The
\emph{slope of $e$} is the angle between $e$ and the $x$-axis, i.e.
$\slope(\Gamma, e) = \arctan(\Delta y / \Delta x)$ if $\Delta x \neq 0$ and
$\slope(\Gamma, e) = -\pi/2$ otherwise.

\subsection{Force-directed Approaches} \label{sec:forces} 

\begin{figure}[t]
	\centering
	\includegraphics[page=4]{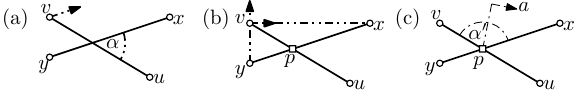}
	\caption{Sketches of the force (a) $F_\mathrm{cos}(v)$, (b)
	$F_\mathrm{cage}(v)$
	and (c) $F_\mathrm{ang}(v)$.}
	\label{fig:forces}
\end{figure}

In general, force-directed algorithms~\cite{SPE:SPE4380211102,10000023432}
compute for each vertex $v$ of a graph $G=(V, E)$ a force $F_v$.  A new drawing
$\Gamma'$ is obtained from a drawing $\Gamma$ by displacing every vertex $v$
according to the force $F_v$.  Classically, the force $F_v$ is a linear
combination of repelling and attracting forces, i.e., all pairs of vertices
repel each other, and incident vertices attract each other. It is easy to
integrate new forces into this generic system, e.g., in order to increase the
crossing angle. For this purpose, Huang et al.~\cite{5635222} introduced the
\emph{cosine force $F_{\cos}$}. The force-directed approach considered by
Argyriou et al.~\cite{DBLP:journals/cj/ArgyriouBS13} uses two forces,
$F_{\mathrm{cage}}$ and $F_{\mathrm{ang}}$, to increase the crossing angle.  In
the following we will describe each force.

\newcommand{\myvec}[1]{\overrightarrow{#1}} Let $\myvec{xy}$ denote the unit
length vector from $x$ to $y$.  Let $uv,xy$ be two crossing edges in $\Gamma$
and let $\alpha$ be the angle as depicted in Fig.~\ref{fig:forces}(a) and let $p$
denote the intersection point of $uv$ and $xy$, see Fig~\ref{fig:forces}. The
cosine force for $v$ is defined as $F_{\cos}(v) = k_{\cos} \cdot\cos \alpha
\cdot \myvec{yx}$, where $k_{\cos}$ is a positive constant.

The force $F_\mathrm{cage}(v)$ is a compound of two forces $F_\mathrm{cage}(v,
x)$ and $ F_\mathrm{cage}(v, y)$.  Let $l_{ab}$ denote the distance between two
points $a$ and $b$.  Let $l^\star_{vx}$ be the length of the edge $vx$ in a
triangle $vxp$ with side length $l_{vp}$ and $l_{xp}$, and a right angle at the
point $p$.  Then, $F_{\mathrm{cage}}(v, x) = k_\mathrm{cage} \cdot \log (
l_{vx} / l^\star_{vx}) \myvec{vx}$, where $k_\mathrm{cage}$ is positive
constant. The force $F_{\mathrm{cage}}(v,y)$ is defined symmetrically.

Again the force $F_\mathrm{ang}(v)$ is a compound of the forces
$F_\mathrm{ang}(v, x)$ and $F_\mathrm{ang}(v, y)$. Consider the unit vector $a$
that is perpendicular to the bisector of $\myvec{uv} $ and $\myvec{yx}$, refer
Fig.~\ref{fig:forces}c. Further, let $\alpha'$ be the angle between the
$\myvec{uv}$ and $\myvec{yx}$. Then the force $F_\mathrm{ang}(v, x)$ is defined
as $k_\mathrm{ang} \cdot \mathrm{sign}(\alpha' - \pi / 2) \cdot | \pi / 2 -
\alpha'| / \alpha' \cdot a$ where $k_\mathrm{ang}$ is a positive
constant. The force $F_\mathrm{ang}(v, y)$ is defined correspondingly.

\section{Multilevel Random Sampling}
\label{sec:random_sampling}

\newcommand{\bestpost}{\mathrm{bestpos}}
\newcommand{\bprating}{\mathrm{rating}}

\begin{algorithm}[bt]
\caption{Random Sampling}
\label{alg:postproc}

	\Input{Initial drawing $\Gamma$, number of levels
		$L \in \N$, number of samples $T \in \N$, scaling factor $b \in (0,1)$, side
	length $s > 0$}
	\Output{Drawing $\Gamma$}	

	\While{\text{stopping criteria}}{
		$(e_1, e_2) \leftarrow$ crossing edges with smallest crossing angle in $\Gamma$\\
		$v \leftarrow$ random vertex in $e_1 \cup e_2$\\
		\For{$i \leftarrow 1$ \KwTo $L$}{
			$R^i \leftarrow $ square centered at $\Gamma[v]$ with side length $s \cdot b^{i-1}$\\
			\For{$1$ \KwTo $T$}{
				$q \leftarrow$ uniform random position in $R^i$\\
				\If{$\Crangle(\moveVertex{v}{q}, v) > \Crangle(\Gamma, v) $}{
					$\Gamma[v] \leftarrow q$\\
				}
			}
		}
	}
\end{algorithm}

Our algorithm starts with a drawing $\Gamma$ of a graph $G$ and iteratively
improves the crossing angle of $\Gamma$ by moving a vertex to a better position,
i.e., we locally optimize the crossing angle of the drawing; for pseudocode
refer to Alg.~\ref{alg:postproc}. For this purpose we greedily select a
vertex~$v$ with a minimal crossing angle $\Crangle(\Gamma, v)$. More precisely,
let $e$ and $e'$ be two edges with a minimal crossing angle in $\Gamma$. We set
$v$ randomly to be an endpoint of $e$ and $e'$. We iteratively improve the
crossing angle of $v$ by sampling a set $S$ of $T$ points within a square $R$
and by moving $v$ to the position $p \in S$ that induces a maximal local
crossing angle. We repeat this process $L \in \N^+$ times and decrease the size
of $R$ in each iteration.

More formally, denote by $\moveVertex{v}{p}$ the drawing obtained from $\Gamma$
by moving $v$ to the point $p=(p_x,p_y) \in \R^2$.  Let $R^i(p) = [p_x - s \cdot
b^i / 2, p_y - s \cdot b^i/2] \times [p_x + s/2, p_y + s \cdot b^i / 2] \subset
\R^2$ be a square centered at the point $p$ with a \emph{scaling factor} $b \in
(0, 1)$ and \emph{initial side length} $s > 0$.  Let $p^0$ be the position of
$v$ in $\Gamma$ and let $S^0 \subset R^0(p^0)$ be a set of $T$ points in
$R^0(p^0)$ chosen uniformly at random. Let $p^i$ be a point in $S^{i-1} \cup
\{p^{i-1}\}$ that maximizes $\Crangle(\moveVertex{v}{p^i}, v)$. We obtain a new
sample $S^i$ by randomly selecting $T$ points within the square~$R^i(p^i)$.
Since $\Crangle(\moveVertex{v}{p^i}, v) =  \max_{uv \in E, e \in E \setminus
\{uv\}} \Crangle(\moveVertex{v}{p^i}, uv, e)$, the function can be evaluated
in $O(\deg(v) |E|)$ time.

\subsection{Fast Minimum Angle Computation}
\label{sec:fast_min_angle_compute}

The running time of the random sampling approach relies on computing in each
iteration a pair of edges creating the minimum crossing angle
$\Crangle(\Gamma)$. More formally, we are looking for a pair of distinct edges
$e, f \in E$ that have a minimal crossing angle in a straight-line drawing
$\Gamma$, i.e., $\Crangle(\Gamma, e, f) = \Crangle(\Gamma)$.  The well known
sweep-line algorithm~\cite{1675432} requires $O( (n+k) \log (n + k))$ time to
report all $k$ intersecting edges in $\Gamma$.  In general the number of
intersecting edges can be $\Omega(m^2)$, but we are only interested in a
single pair that forms the minimal crossing angle.  Therefore, we propose an
algorithm, which uses the slopes of the edges in $\Gamma$ to rule out pairs of
edges, which cannot form the minimum angle.

Assume that we already found two intersecting edges forming a small angle of
size $\delta > 0$.  We set $t := \left\lfloor \pi / \delta \right\rfloor$ and
distribute the edges into $t$ buckets $B_0, \ldots, B_{t-1}$ such that bucket
$B_i$ contains exactly the edges $e$ with $i\pi/t \leq \slope(\Gamma, e) + \pi/2
< (i+1)\pi/t$.  Then each bucket covers an interval of size $\pi / \left\lfloor
\pi / \delta \right\rfloor \geq \delta$.  Thus, if there exist edges $e, f$ with
$\Crangle(\Gamma, e, f) < \delta$, they belong to the same or to the adjacent
buckets (modulo $t$). Overall, we consider all pairs of edges in $B_i \cup
B_{i+1\Mod t}$, $i=1,\dots t$, and find the pair forming the smallest crossing
angle.
%
%
To find this pair  we could apply a
sweep-line algorithm to the set $B_i \cup  B_{i+1}$.  In general this set can contain $\Omega(m)$ edges. 
Thus, in worst case we would not gain a
speed up in comparison to a sweep-line algorithm applied to $\Gamma$.  On the
other hand, in practice we expect the number of edges in a bucket to be small.  If
we assume this number to be a constant, the overall running time of the exhaustive check 
is linear in $m$ and does not depend on the number of
crossings.

\myparagraph{Implementation Details} In the case that the slopes in $\Gamma$ are
uniformly distributed, we expect the number of edges in a bucket to decrease
with an decreasing estimate $\delta$.  We set the value $\delta$ to be the
minimal crossing angle of the $r$ longest edges in $\Gamma$. In our
implementation we set $r$ to be $50$ if the graph contains at most $5000$ edges,
otherwise it is $300$. 

\section{Experimental Evaluation}
\label{sec:evalution}

The \randomSampling heuristic has several parameters which allow for many
different configurations. In Sec.~\ref{sec:eval:random_sampling}, we investigate
the influence of the configuration on the crossing angle of the drawing computed
by the random sampling approach. 
We investigate the question of whether the \randomSampling approach improves the
crossing angle of a given drawing. Our evaluation in
Sec.~\ref{sec:eval:improvement} answers the question affirmatively.  Moreover,
we expect that the crossing angle of the drawing computed by the random sampling
approach depends on the choice of the initial drawing.  We show that this is
indeed the case (Sec.~\ref{sec:eval:effect_id}). 
We close the evaluation with a short running time analysis in
Sec.~\ref{sec:running_time}.  Our evaluation is based on a selection of
artificial and real world graphs (Sec.~\ref{sec:graph_classes}), several choices
of the initial drawing, see Sec.~\ref{sec:initial_drawings}, and a specific way
to compare two drawing algorithms (Sec~\ref{sec:binomial_adv_test}).

\myparagraph{Setup}
All experiments were conducted on a single core of an AMD Opteron Processor 6172
clocked at 2.1~GHz. The server is equipped with 256GB RAM. All algorithms were
compiled with \texttt{g++-4.8.5} with optimization mode \texttt{-O3}.

\subsection{Benchmark Graphs}
\label{sec:graph_classes}

\begin{figure}[tb]
	\centering
	\subfloat[\north]{
		\includegraphics{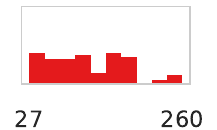}
	}
	\subfloat[\rome]{
		\includegraphics{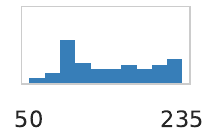}
	}
	\subfloat[\community]{
		\includegraphics{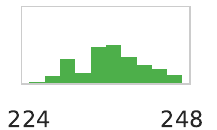}
	}
	\subfloat[\Oneplanar]{
		\includegraphics{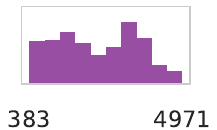}
	}
	\subfloat[\textsc{Tri.+X}]{
		\includegraphics{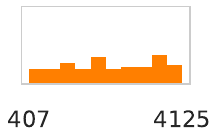}
	}
	\caption{The distribution of the sum of number of vertices and edges per graph
	class. The plot is scaled such that a bar of full height would contain 40
graphs.}
	\label{fig:size_distribution}
\end{figure}

We evaluate the heuristic on the following graph classes, either purely
synthetic or with a structure resembling real-world data.  
Fig.~\ref{fig:size_distribution} shows the size
distribution of these graphs. The color of each class is used consistently
throughout the paper.

\myparagraph{Real World} The classes \emph{Rome} and \emph{North}
(AT\&T)\footnote{http://graphdrawing.org/data.html} are the non-planar subsets
of the corresponding well known benchmark sets, respectively.  From each graph
class we picked 100 graphs uniformly at random.  The \community graphs are
generated with the \textsc{LFR-Generator}~\cite{PhysRevE.78.046110} implemented
in \textsc{NetworKit}~\cite{DBLP:journals/netsci/StaudtSM16}. These graphs
resemble social networks with a community structure.  

\myparagraph{Artificial} For each artificial graph we picked the number $n$ of
vertices uniformly at random between 100 and 1000.  The \TriX class contains
randomly generated $n$-vertex triangulations with an additional set of $x$
edges. The number $x$ is picked uniformly at random between $0.1n$ and
$0.15n$. The endpoints of the additional edge are picked uniformly at
random, as well.

The class \Oneplanar consists of graphs that admit drawings where every edge has
at most one crossing. We used a \emph{geometric} and \emph{topological}
procedure to generate these graphs. For the former consider a random point set
$P$ of $n$ points. 
Let $e_1, \dots e_k$ be a random permutation of all pairs of points in $P$.
Let $G_0 = (P, \emptyset)$. If the drawing $G_{i-1} + e_i$ induced by $P$ is
simple and $1$-planar, we define $G_i$ to be this graph, otherwise we set $G_i =
G_{i-1}$.
We construct the topological \Oneplanar graphs based on a random planar
triangulation $G$ generated with OGDF~\cite{ogdf}.  Let $v$ be a random vertex
of $G$ and let $v, x, u, y$ be an arbitrary $4$-cycle. We add $uv$ to $G$ if
$G+uv$ is 1-planar. The process is repeated $x$ times, for a random number $x
\in [0.3n, 0.4n]$.
In contrast to the experimental work on crossing minimization in book
embeddings~\cite{DBLP:conf/gd/KlawitterMN17}, we did not observe that our
heuristic performs differently on the topological and geometric \Oneplanar
graphs\iflong
, see Appendix~\ref{sec:1planar}
\fi.
Hence, we merge the two classes into a
single class.  Thus, in total the \Oneplanar graphs contain 200 graphs compared
to 100 in the other graph classes.

\subsection{Initial Drawings}
\label{sec:initial_drawings}
\begin{figure}[tb]
	\centering
	\includegraphics{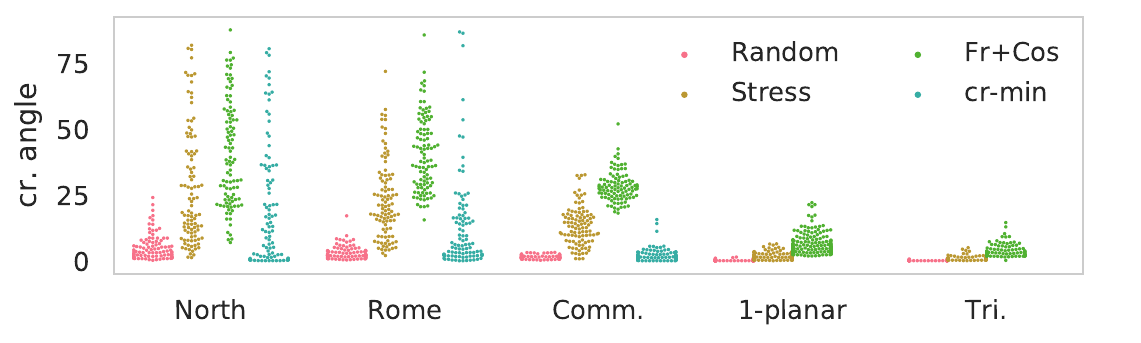}
	\caption{Crossing angles of the initial drawings.}
	\label{fig:initial_angles}
\end{figure}

In our evaluation we consider four initial drawings of each benchmark graph;
refer to Tab.~\ref{tab:short_names_algorithms}. A random point set $P$ of size
$n$ induces a \rnd drawing of an $n$-vertex graph. 
The \frcos drawings are generated by applying our implementation of the
force-directed method of Fruchtermann and Reingold~\cite{SPE:SPE4380211102} to
the \rnd drawings with the additional $F_\mathrm{cos}$ force
(Sec.~\ref{sec:forces}).
We applied the stress majorization~\cite{ogdf,Gansner2005} implementation of the
\textsc{Open Graph Drawing Framework} (OGDF) to \rnd in order to obtain the
\stress drawings.
The \crmin drawings are computed with the heuristic introduced by Radermacher et
al.~\cite{doi:10.1137/1.9781611975055.12} in order to decrease the number of
crossings in straight-line drawings. They showed that the heuristic computes
drawings with significantly less crossings than drawings computed by stress
majorization.  Unfortunately, within a feasible amount of time we were not able
to compute \crmin drawing for the classes \Oneplanar and \TriX.

A point in Fig.~\ref{fig:initial_angles} corresponds to the crossing angle of an
initial drawing. The plot is categorized by graph class. The \rnd drawings have
the smallest crossing angles. The  \stress drawings have a larger crossing angle
than \crmin and overall,  \frcos drawings tend to have the largest crossing
angle. 

We point out that in contrast to the evaluation of Argyriou et
al.~\cite{DBLP:journals/cj/ArgyriouBS13}, our implementation of the
force-directed method with $F_{\mathrm{cage}}$ and $F_{\mathrm{ang}}$ produces
drawings with smaller crossing angles than with $F_\mathrm{cos}$. Thus, we do
not consider these drawings in our evaluation.
\iflong For the further details refer to Appendix~\ref{sec:cage_angular_force}.
\fi

\begin{table}[tb]
	\centering
	\subfloat[\label{tab:short_names_algorithms}]{
	\begin{tabular}{l l}
		\textbf{Identifier} & \textbf{Algorithm} \\
		\hline \rnd & uni. rand. vertex
		placement\\ \frcos & FR  +  Cosine Forces (Sec.~\ref{sec:forces})\\
		\stress & Stress Majorization~\cite{Gansner2005} \\
		\crmin & Crossing Minimization~\cite{doi:10.1137/1.9781611975055.12}\\
	\end{tabular}
 	}
	\quad\quad
	\subfloat[\label{tab:post}]{
	\begin{tabular}{lccccc}
		& \textbf{Levels} &\quad & \textbf{Sample Size} \\ 
		& $L$		& & $T$		\\
		\hline \Sloppy  & 3 && 50  \\
		\Medium & 4 && 175 \\%
		\Precise & 5 && 400 \\
	\end{tabular}
}

 \caption{ (a) Initial drawings with their identifiers used throughout the
 paper.  (b) Configurations of the \randomSampling approach. The scaling
 factor $b$ is $0.2$ and the initial side length $s$ is
 $10^5$.  }

\end{table}

\subsection{Differences between Paired Drawings}
\label{sec:binomial_adv_test}

In order to compare the performance of two algorithms on multiple graphs and to
investigate by how much one of the algorithms outperforms the other, we employ
the following machinery.  We denote by $\Gamma\{G\}$ the set of all drawings of~$G$.  Let $\mathcal{G} = \{G_1, G_2, \dots, G_k\}$ be a family of (non-planar)
graphs.  We refer to a set $\Lambda = \{\Gamma_1, \dots, \Gamma_k\}$ as a
\emph{family of drawings} of $\mathcal G$ where $\Gamma_i \in \Gamma\{G_{i}\}$ .
Let $\Lambda^1$ and $\Lambda^2$ be two families of drawings of $\mathcal G$.
Let $\mathcal F$ be a subset of $\mathcal G$. We say \emph{$\Lambda^1$
outperforms $\Lambda^2$ on $\mathcal F$} if and only if for all $G_i \in
\mathcal F$ the inequality $\Crangle(\Gamma_i^1) > \Crangle(\Gamma^2_i)$ holds.
If $\Lambda^1$ outperforms $\Lambda^2$ on $\mathcal F$ then $\Lambda^1$ has an
\emph{advantage of $\Delta > 0$ on $\mathcal F$} if for all $G_i \in \mathcal F$
the inequality $\Crangle(\Gamma^1_i) > \Crangle(\Gamma^2_i) + \Delta$ holds.
For a finite set $\mathcal G$, we say $\mathcal F$ has \emph{relative size at
least $p\in[0,1]$} if $|\mathcal F| \geq p \cdot |\mathcal G|$. 

In order to compare two families of drawings we plot the advantage as a function
of $p$; refer to Fig.~\ref{fig:random_sampling:test}. For each value $p$ the
plot contains 5 five bars, each corresponding to a graph class. The height of
the bars correspond to advantages $\Delta$ for a set of relative size $p$.  A
caption of a figure in the form of \emph{A vs B} indicates that if $\Delta$ is
positive, $B$ has advantage $\Delta$ over A. Correspondingly, if $\Delta$ is
negative, $A$ has an advantage of $-\Delta$ over $B$.  Thus,
Fig.~\ref{fig:random_sampling:test} shows that for $p=0.1$, for each graph class
there is a subset $\mathcal F$ of relative size $0.1$, i.e., $\mathcal F$
contains at least 10 graphs, such that the set \Sloppy has an advantage of
$\Delta$ over \Precise on $\mathcal F$. In greater detail, \Sloppy has an
advantage of $7.9^\circ$ over \Precise on the \north graphs, $12.9^\circ$ on the
\rome graphs, $11.5^\circ$ on the \community graphs, $1.2^\circ$ on the
\Oneplanar graphs and $1.2^\circ$ on the \TriX graphs.  On the other side,
\Precise has an advantage of $12.9^\circ$ over $\Sloppy$ on at least 10 \north
graphs, $15.7^\circ$ on the \rome graphs, $13.8^\circ$ on the \community graphs,
$1.1^\circ$ on the \Oneplanar graphs and $0.4^\circ$ on the \TriX graphs.  Note
that only for $p < 0.5$ there can be two disjoint subsets $\mathcal F_1,
\mathcal F_2$ of a graph class of relative size $p$  such that \Precise has an
advantage over \Sloppy on $\mathcal F_1$ and \Sloppy has an advantage over
\Precise on $\mathcal F_2$. 

\subsection{Parametrization of the Random Sampling Approach}
\label{sec:eval:random_sampling}

The \randomSampling approach introduced in Section~\ref{sec:random_sampling} has
four different parameters, the number of levels $L$, the size of the sample $T$,
the initial side length $s$ and the scaling factor $b$, that allows for many
different configurations. With an increasing number $T$ of samples, we expect to
obtain a larger crossing angle in each iteration to the cost of an increasing
running time. If we allow each configuration the same running time, it is
unclear whether it is beneficial to increase the number of iterations or to
increase the number of samples ($T$) and levels ($L$) per iteration.  This
motivates the following question: does the crossing angle of a drawing of an
$n$-vertex graph computed by the random sampling approach within a given time
limit $t_n$ increase with an increasing number of samples and levels?
\begin{figure}[tb]
	\centering
	\includegraphics{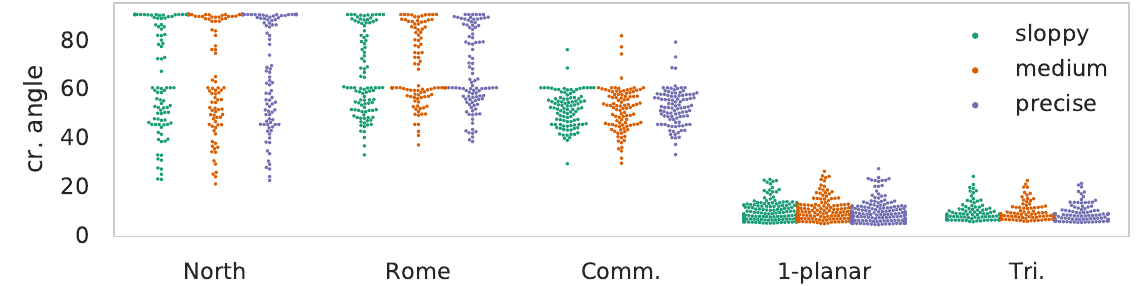}
	\caption{Performance of different configurations}
	\label{fig:sampler_performance}
\end{figure}
We choose to set the time limit $t_n$ to $n$ seconds.  This allows for at least
$1.6 \cdot n$ iterations on our benchmark instances.
\iflong
For further details see Appendix~\ref{apx:nof_iterations}.
\fi
Since the parametrization space is
infeasibly large, we evaluate three exemplary configurations, \Sloppy, \Medium
and \Precise; see Tab.~\ref{tab:post}. 

\begin{figure}[tb] \centering
	
	\subfloat[\Precise vs \Sloppy]{
		\includegraphics{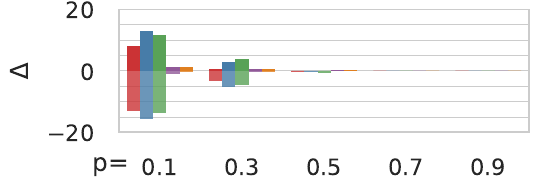}
	}
	\quad
	\subfloat[\Precise vs \Medium]{
		\includegraphics{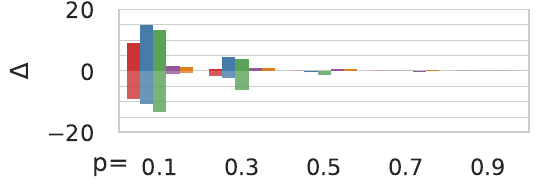}
	}
	
	\caption{Comparison of the \Sloppy configuration to the \Medium and \Precise
		configuration. The colors indicate the graph as indicated by
		Fig.~\ref{fig:size_distribution}.} \label{fig:random_sampling:test}

\end{figure}

The plot in Fig.~\ref{fig:sampler_performance} does not indicate that the
distributions of the crossing angle differ across different configurations
significantly
\iflong
; further characteristics are listed in Tab.~\ref{tab:char:configs}
in the appendix
\fi.
With the plot in Fig.~\ref{fig:random_sampling:test} we can
confirm this observation.  For each configuration there is only a small subset
of each class such that the configuration has an advantage over the other
configurations. For example, for the \rome graphs there exist at least 10 graphs
such that \Sloppy has an advantage of $10^\circ$ over \Precise. On the other
hand, there are at least 10 different graphs such \Precise has also an advantage
of $10^\circ$ over \Sloppy.  For $p\geq 0.5$ no configuration has an advantage
over the other, or it negligibly small.  Thus, we conclude that given a common
time limit, increasing the levels and the sample size does not necessarily
increase the crossing angle.  

\subsection{Improvement of the Crossing Angles} 
\label{sec:eval:improvement}

In this section we investigate whether the \randomSampling approach is able to
improve the crossing angle of a given drawing within $2n$ iterations.  Given the
same number of iterations, it is most-likely that we obtain a larger crossing
angle of a drawing if we increase the number of samples. Thus, we use the
\Precise configuration for the evaluation of the above question.  We refer to
the drawings after the application of the \randomSampling approach 
as \rndRS, \frcosRS, \stressRS and \crminRS, respectively. 
\iflong
For characteristics
of the crossing angles refer to Tab.~\ref{tabl:summary} in the appendix.
\fi

\begin{figure}[tb]
	\centering
		\includegraphics{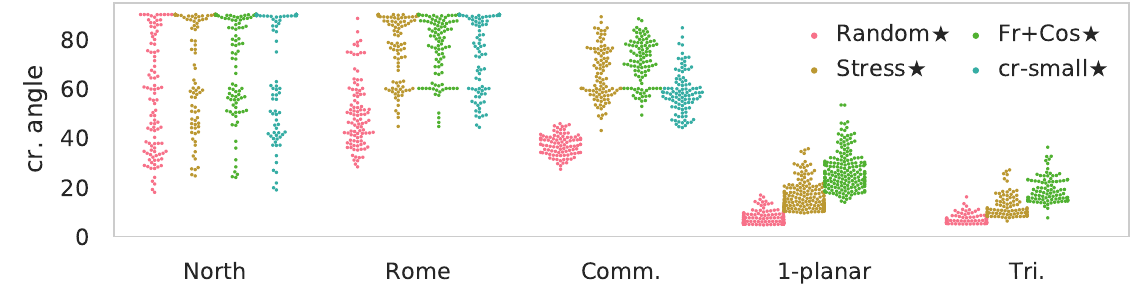}
	\caption{Crossing angles of
		the initial drawings after optimization with the \randomSampling approach. 
	}
	\label{fig:initial_layout:final_angles}
\end{figure}

\begin{figure}[tb]
	\centering
	\subfloat[\rnd]{
		\includegraphics{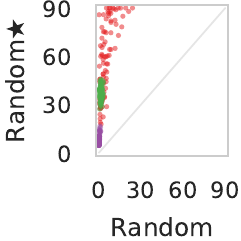}
	}
	\subfloat[\crmin]{
		\includegraphics{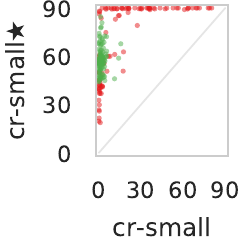}
	}
	\subfloat[\stress]{
		\includegraphics{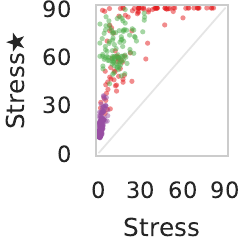}
	}
	\subfloat[\frcos]{
		\includegraphics{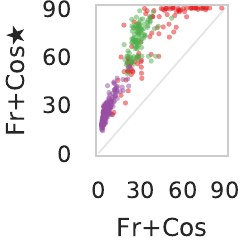}
	}
	\caption{Initial crossing angle vs the final crossing angle. The plots show
		the crossing angles of the classes \north, \community and \Oneplanar.}
	\label{fig:ivf:direct}
\end{figure}

The plots in Fig.~\ref{fig:initial_angles} and
Fig.~\ref{fig:initial_layout:final_angles} indicate that the \randomSampling
approach indeed improves the crossing angle of the initial drawings.
Fig.~\ref{fig:ivf:direct} shows the relationship between the crossing angle of
the initial drawing and the final drawing. For the purpose of clarity, the plot
only shows drawings of the classes \north, \community and \Oneplanar
\iflong
; for the
remaining classes refer to Appendix~\ref{apx:improvement}
\fi.
The plots shows
that the \randomSampling approach considerably improves the crossing angle of
the initial drawing.  In case of the \north graphs there are a few graphs that
have an improvement of at least $70^\circ$.  There are at least 10 drawings in
\rnd whose crossing angle is improved by at least $75^\circ$
\iflong
; refer to
Fig.~\ref{fig:ivf} in Appendix~\ref{apx:improvement}\fi.
For all real
world graph classes and all initial layouts there are $70$ graphs in each class,
such that the final drawing has an advantage of over $25^\circ$.

For \TriX, \frcosRS has an advantage of at least $11^\circ$ over \frcos on
at least 90 \TriX. For the remaining initial layouts the corresponding advantage
is at most $7.6^\circ$.  Considering the \Oneplanar graphs the corresponding
advantages are $14^\circ$ and $9.7^\circ$.  This indicates that within $2n$
iterations a large initial crossing angle helps to further improve the crossing
angle of \Oneplanar and \TriX graphs.  Overall we observe that the \Oneplanar
and  \TriX classes are rather difficult to optimize.  This can either be a
limitation of our heuristic or the crossing angle of these graphs are indeed
small.  Unfortunately, we are not aware of meaningful upper and lower bounds on
the crossing angle of straight-line drawing of these graphs.  Nevertheless, we
can conclude that our heuristic indeed improves the initial crossing angle. To
which extend our heuristic is able to increase crossing angle of a drawing
depends on the graph class and on the initial drawing itself.

\subsection{Effect of the Initial Drawing}
\label{sec:eval:effect_id}

The \randomSampling approach iteratively improves the crossing angle of a given
drawing. Given a different drawing of the same graph the heuristic might be able
to compute a drawing with a larger crossing angle. Hence, we investigate
whether the choice of the initial drawing influences the crossing angle of a
drawing obtained  by the \randomSampling approach with $2n$ iterations.

For all graph classes, except from \north, it is apparent from
Fig.~\ref{fig:initial_layout:final_angles} that the drawings in the set \rndRS
have noticeably smaller crossing angles compared to the remaining drawings. This
meets our expectations, since the initial \rnd drawings presumably has many
crossings~\cite{Huang:2010:ERI:1865841.1865854} and thus is likely to have many small crossing angles; compare the
initial crossing angles plotted in Fig.~\ref{fig:initial_angles}.

\begin{figure}[bt]
	\centering
	\subfloat[\rnd \label{fig:initial_layout:paired_plot:rnd}]{
		\includegraphics{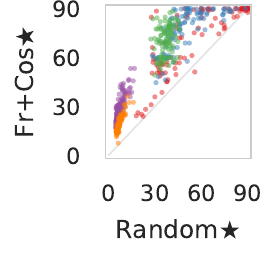}
	}
	\quad
	\subfloat[\crmin \label{fig:initial_layout:paired_plot:crmin}]{
		\includegraphics{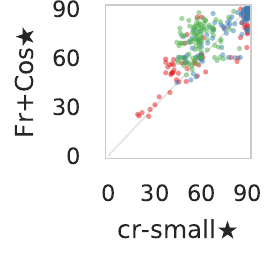}
	}
	\quad
	\subfloat[\stress \label{fig:initial_layout:paired_plot:stress}]{
		\includegraphics{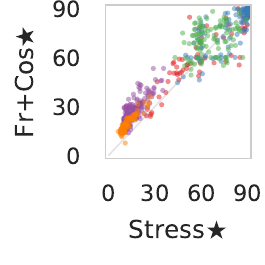}
	}
	\caption{Comparison of the initial layout.}
	\label{fig:initial_layout:paired_plot}
\end{figure}

The plot in Fig.~\ref{fig:initial_layout:final_angles}
\iflong
and the characteristics
in Tab.~\ref{tabl:summary} in the Appendix suggest
\else 
suggests
\fi
that the set \frcosRS
contains drawings with the largest crossing angles. In order to corroborate this
claim, Fig.~\ref{fig:initial_layout:paired_plot} shows crossing angles obtained
by different algorithms.
%
%
It shows that except for one graph, each drawing in \frcosRS has a larger
crossing angle than the corresponding drawing in \rndRS.
Fig~\ref{fig:initial_layout:paired_plot:crmin} and
Fig.~\ref{fig:initial_layout:paired_plot:stress} suggest that \frcosRS overall
contains more drawings with a larger crossing angle compared to \stressRS and
\crminRS. 
\begin{figure}[bt] \centering \subfloat[\stressRS vs \frcosRS
	\label{fig:initial_layout:frcos:stress}]{
		\includegraphics{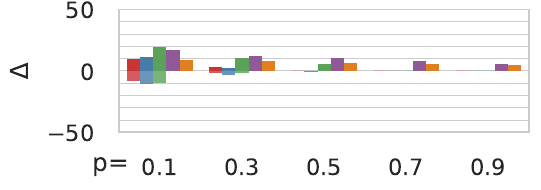}
	} \quad \subfloat[\crminRS vs \frcosRS \label{fig:initial_layout:frcos:cr}]{
		\includegraphics{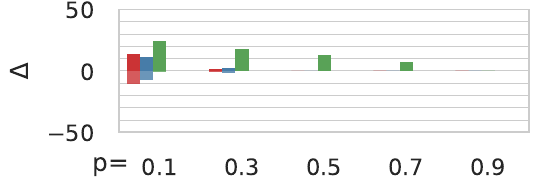}
	}

	\subfloat[\rndRS vs \frcosRS \label{fig:initial_layout:frcos:rnd}]{
		\includegraphics{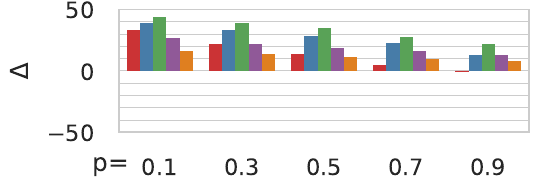}
	} \quad \subfloat[\stressRS vs \crminRS \label{fig:initial_layout:cr:stress}]{
		\includegraphics{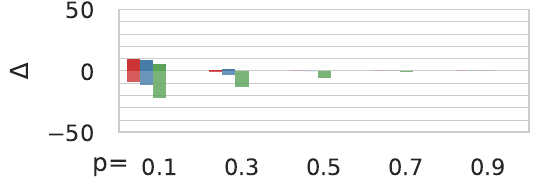}
	} \caption{Comparison of the crossing angle of the final drawings.}
	\label{fig:initial_layout:frcos} \end{figure}
With the help of Fig.~\ref{fig:initial_layout:frcos} we are able to quantify the
number of graphs above the diagonal and the difference of the crossing angles.
Fig.~\ref{fig:initial_layout:frcos:stress} shows that for the graph classes
\Oneplanar and \TriX, there are each at least 90 of 100 graphs whose drawings in
\frcosRS have a crossing angle larger then the corresponding drawing in
\stressRS, i.e., \frcosRS has an advantage of $4.5^\circ$ degrees over \stressRS.

There are at least 50 \Oneplanar graphs such that the \frcosRS has an advantage
of $10^\circ$ over \stressRS.  At least 50 \community graphs have
drawings in \frcosRS with an advantage of $5^\circ$ over the corresponding
drawings in \stressRS.  There are 10 \north graphs such that \frcosRS  has an
advantage of at least $5^\circ$ over \stressRS. Vice versa there are 10
different \north graph such that \stressRS has an advantage of at least $5^\circ$
degrees over \frcosRS.  Considering subsets of size $10$, \frcosRS has an
advantage of $20^\circ$ over \stressRS.

Recall that \crminRS does neither contain  drawings of the class \Oneplanar nor
of the class \TriX.  The drawings of \frcosRS has an advantage of over $7^\circ$
over \crminRS on over 70 \community graphs. For a subset with at least 10
\community graphs, the advantage rises to almost~$25^\circ$.
The comparison on \stressRS and \crminRS shows that drawings with a few
crossings do not necessarily yield larger crossing angles.  Overall, we conclude
that the \randomSampling approach computes the largest crossing angle when
applied to the \frcos drawings.  This is plausible, since the crossing angles of
the initial crossing angles are already good. As shown in the previous section,
depending on the graph class, there is a large improvement in the crossing
angle, if we start with such an initial drawing. 
\iflong
In
Appendix~\ref{sec:4n_Iterations}  we show that the advantages of \frcosRS
decreases comparably to \stressRS with $4n$ iterations.
\else
In further investigations we were able to show that the advantages of \frcosRS
decreases comparably to \stressRS with $4n$ iterations.
\fi
However, doubling the
iterations does not entirely cover the gap between the crossing angles of the
initial drawings.


\subsection{Note on the Running Time}
\label{sec:running_time}

\begin{figure}[tb]
	\centering
	\includegraphics{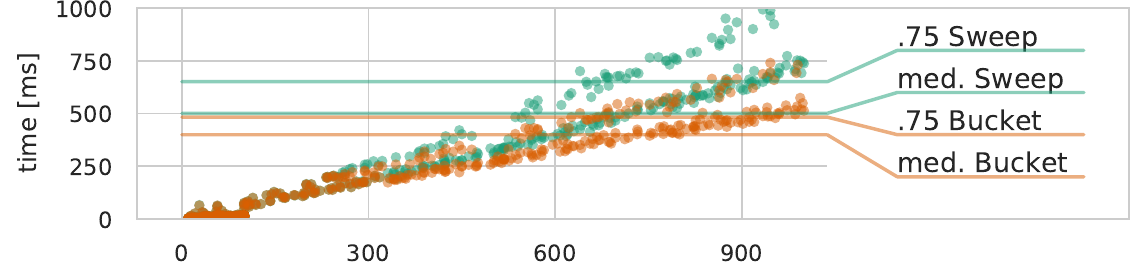}
	\caption{Average Running time per iteration vs the number of vertices.}
	\label{fig:slow_fast}
\end{figure}

In this section we shortly evaluate the running time of our algorithm on all our
graphs. For this purpose, we applied two implementations of the \randomSampling
heuristic to the \rnd drawings. The \textsc{Sweep} implementation uses a
sweep-line algorithm to compute the pair of crossing edges that create the
smallest crossing.  \textsc{Bucket} uses the algorithm described in
Sec.~\ref{sec:fast_min_angle_compute}. We employ the speed-up technique only for
graphs with at least 1000 edges, we refer to these graphs as \emph{large}.
Fig~\ref{fig:slow_fast} plots the running time per iteration for $n$-vertex
graphs. The median and the second 3-quantile of the running time on the large
graphs are highlighted. \textsc{Bucket} has an average running time of 391ms per
iteration on the large graphs and \textsc{Sweep} has an average running time of
500ms. On all graph \textsc{Bucket} requires on average 328ms per iteration. 
\iflong
In Appendix~\ref{sec:cr_vs_ops}, we compare the number of crossings with the
number of tested edge-pairs.
\fi

\section{Conclusion}

\begin{figure}[b]
	\centering
	\subfloat[]{
		\includegraphics[height=0.15\textwidth]{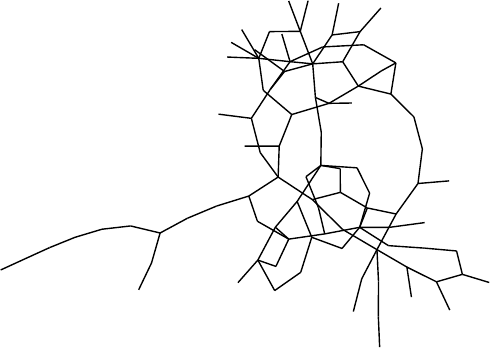}
	}
	\quad\quad	
	\subfloat[]{
		\includegraphics[page=2, height=0.15\textwidth]{fig/joined.pdf}
	}

	\caption{(a) \stress drawing of a Rome graph. (b) Drawing after optimizing the
	crossing angle. The ratio between the longest and shortest edges is large.}
	\label{fig:examples}
\end{figure}

We designed and evaluated a simple heuristic to increase the crossing angle in a
straight-line drawing of a graph. On real world networks our heuristic is able
to compute larger crossing angles than on artificial networks. This can either
be a limitation of our heuristic or the crossing angle of our artificial graph
classes are small.  We are not aware of lower and upper bounds of the crossing
angle of these graphs. Thus, investigating such bounds of the \Oneplanar and
\TriX graphs is an interesting theoretical question.

Fig.~\ref{fig:examples} shows that our heuristic does not necessarily compute
readable drawings. Nevertheless, parts of the \randomSampling heuristic are
easily exchangeable. For example, the objective function can be replaced by a
linear combination of number of crossing and the crossing angle. 
Thus, future work can be concerned with adapting the \randomSampling approach
with the aim to compute readable drawings.


\bibliographystyle{splncs04}
\bibliography{strings,references}

\iflong
\clearpage
\appendix

\section{1-Planar Graphs}

\label{sec:1planar} Fig.~\ref{fig:1planar} shows the crossing angle for the
geometric and topological 1-planar graphs computed with the random sampling
approach. The plot already suggest that the distributions do not differ to much.
Indeed, a t-test confirms that there is no significant difference between the median
of the two distributions.

\begin{figure}
	\centering
	\includegraphics{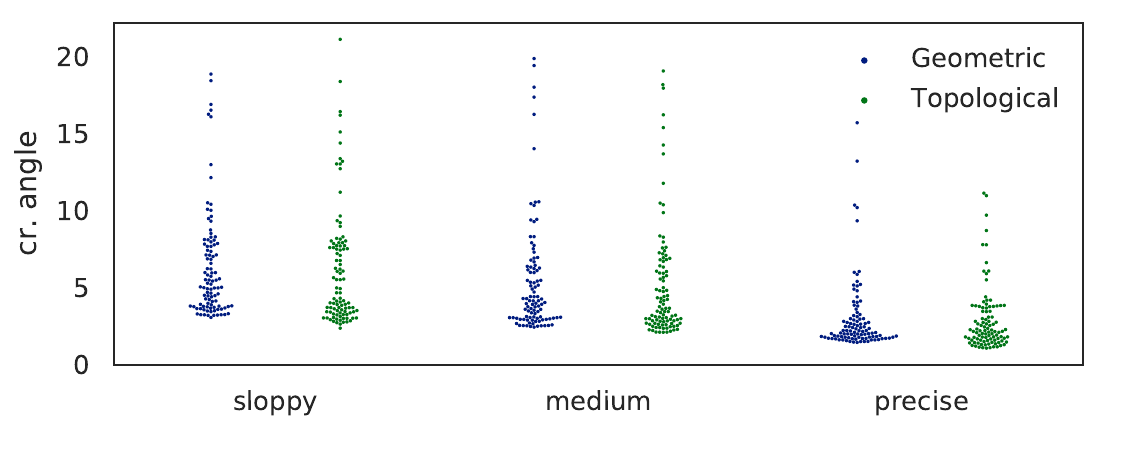}
	\caption{Results for the two 1-planar graph classes}
	\label{fig:1planar}
\end{figure}

\clearpage

\section{Cage and Angular Force}
\label{sec:cage_angular_force}

In this section, we evaluate the crossing angle of drawings computed by
force-directed methods applied to a random initial drawing.  Stress and FMMM are
implementations provided by the OGDF library. The \emph{cage} drawings are
obtained by the force-directed method of Fruchtermann and
Reingold~\cite{SPE:SPE4380211102} with the additional forces $F_{\mathrm{cage}}$
and $F_{\mathrm{ang}}$.  The drawings in Fr+Cos use the $F_\emph{cos}$ force
instead of $F_{\mathrm{cage}}$ and $F_{\mathrm{ang}}$.  We use our own
implementation of the  of the force-directed method of Fruchtermann and
Reingold.

The plot in Fig.~\ref{fig:force_directed} shows the final crossing angles. It
indicates that the cage force produces drawings with the smallest crossing
angles. This is not in accord with the claim of Argyriou et
al.~\cite{DBLP:journals/cj/ArgyriouBS13} that they obtained drawings with the largest
crossing angle using their implementation of the forces $F_\mathrm{cage}$ and
$F_{\mathrm{ang}}$. Our results are not necessarily comparable, since we may
have used different constants to scale the forces. Moreover, we start from different
initial drawings. We always start with a random drawing where Argyriou et al.
use an organic layout (\texttt{SmartOrganic}) provided by yEd (www.yworks.com). 
		
\begin{figure}[h]
	\centering
	\includegraphics{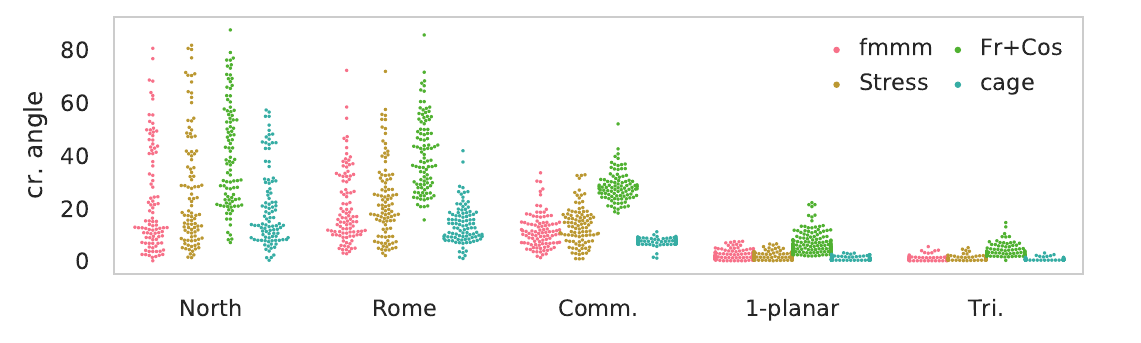}
	\caption{Comparing Force Directed Methods}
	\label{fig:force_directed}
\end{figure}

\section{Characteristics}

Tab.~\ref{tab:char:configs} shows several characteristics of crossing angles
computed with the \Sloppy, \Medium and \Precise configuration of the random
sampling approach. Tab.~\ref{tabl:summary} shows characteristics of the crossing
angle after the application of the \Precise configuration to different initial
drawings.

\begin{table}
	\centering
\caption{Characteristics computed by different configurations.}
\label{tab:char:configs}

\begin{tabular}{llrrrr}
	\textbf{graph class} & \textbf{algorithm} & \multicolumn{4}{c}{
		\textbf{crossing resolution}} \\
	&								& \textbf{min}&  \textbf{mean} & \textbf{median} & \textbf{max} \\
	 \hline
	    Comm. &    \Medium &               29.13 &  51.14 &  52.23 &  81.27 \\
      Comm. &   \Precise &               32.63 &  52.01 &  52.07 &  78.70 \\
      Comm. &    \Sloppy &               28.90 &  51.12 &  51.66 &  75.61 \\
			\hline
      North &    \Medium &               20.63 &  67.12 &  63.87 &  90.00 \\
      North &   \Precise &               22.06 &  67.82 &  68.69 &  90.00 \\
      North &    \Sloppy &               22.49 &  65.84 &  60.00 &  90.00 \\
			\hline
       Rome &    \Medium &               36.58 &  67.85 &  60.00 &  90.00 \\
       Rome &   \Precise &               37.97 &  66.43 &  60.00 &  90.00 \\
       Rome &    \Sloppy &               32.52 &  64.86 &  59.98 &  90.00 \\
			 \hline
			1-planar &    \Medium &                4.33 &   9.02 &   7.36 &  25.79 \\
   		1-planar &   \Precise &                3.92 &   8.60 &   6.97 &  26.84 \\
  		1-planar &    \Sloppy &                4.58 &   8.71 &   7.23 &  22.50 \\
				\hline
       Tri. &    \Medium &                5.27 &   8.94 &   7.66 &  22.03 \\
       Tri. &   \Precise &                4.90 &   8.55 &   7.58 &  20.88 \\
       Tri. &    \Sloppy &                5.13 &   8.90 &   7.55 &  23.71 \\
     \end{tabular}
\end{table}

\begin{table}
	\centering
\caption{Characteristics}
\label{tabl:summary}

\begin{tabular}{llrrrr}
	\textbf{graph class}&  \textbf{layout}& \multicolumn{4}{c}{\textbf{crossing
	resolution}} \\
	&								& \textbf{min}&  \textbf{mean} & \textbf{median} & \textbf{max} \\
\hline
      Comm. &  \frcosRS &               49.16 &  70.63 &  71.10 &  88.25 \\
      Comm. &  \rndRS &               27.18 &  37.09 &  37.24 &  45.68 \\
      Comm. &  \crminRS &               44.09 &  58.54 &  58.03 &  84.61 \\
      Comm. &  \stressRS &               42.89 &  65.91 &  63.75 &  89.09 \\
			\hline
      North &  \frcosRS &               23.82 &  71.29 &  78.83 &  90.00 \\
      North &  \rndRS &               17.81 &  55.87 &  54.51 &  90.00 \\
      North &  \crminRS &               18.77 &  70.55 &  87.87 &  90.00 \\
      North &   \stressRS &               24.46 &  70.84 &  84.68 &  90.00 \\
			\hline
       Rome &  \frcosRS &               44.52 &  77.16 &  81.28 &  90.00 \\
       Rome &  \rndRS &               28.14 &  49.94 &  47.25 &  88.43 \\
       Rome &  \crminRS &               44.19 &  76.32 &  84.28 &  90.00 \\
       Rome &  \stressRS &               44.55 &  77.09 &  82.70 &  90.00 \\
			 \hline
   1-planar &     \frcosRS &               13.76 &  26.55 &  25.25 &  53.26 \\
   1-planar &    \rndRS &                4.55 &   6.91 &   6.02 &  16.67 \\
   1-planar &          \stressRS &                9.38 &  15.81 &  13.85 &  35.50 \\
			\hline
       Tri. &     \frcosRS &                7.43 &  18.77 &  17.24 &  36.13 \\
       Tri. &    \rndRS &                4.92 &   6.79 &   6.20 &  15.94 \\
       Tri. &          \stressRS &                6.14 &  11.95 &  10.41 &  26.89 \\

\end{tabular}
\end{table}
\clearpage

\clearpage
\section{Number of Iterations}
\label{apx:nof_iterations}

Fig.~\ref{fig:n_vs_itr} compares the number of vertices of a graph with the
number of iterations done in the given time limit. The color indicate the
configuration; green is \Sloppy, orange is \Medium, purple is \Precise. It
shows that independent of the configuration,  on each $n$-vertex graph at least
$n$ iterations have been performed.
The actual number of iterations depend on
the configuration. Especially, the \Sloppy configuration is able to do at least
$10n$ iterations on most graphs. On the other hand, \Precise moves considerably
less vertices.

\begin{figure}[h]
	\centering
	\includegraphics{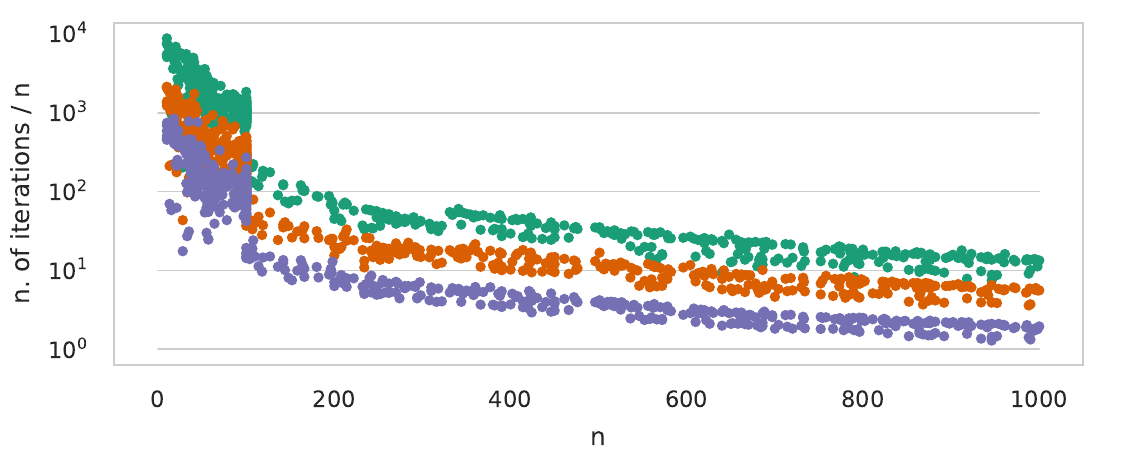}
	\caption{Number $n$ of vertices vs. number of iterations.}
	\label{fig:n_vs_itr}
\end{figure}

\clearpage
\section{Improvement of the Crossing Angle}
\label{apx:improvement}

Fig.~\ref{fig:ivf:direct_apx} shows the missing data of the plot in
Fig.~\ref{fig:ivf:direct}, i.e., the crossing angles of the graph classes \rome
and \TriX. Fig.~\ref{fig:ivf} shows the corresponding advantages. The plots
confirm that our approach significantly increases the crossing angles of the in
the initial drawings.

\begin{figure}
	\centering
	\subfloat[\rnd]{
		\includegraphics{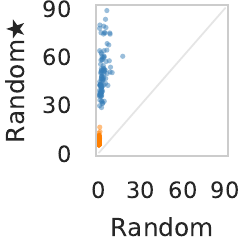}
	}
	\subfloat[\crmin]{
		\includegraphics{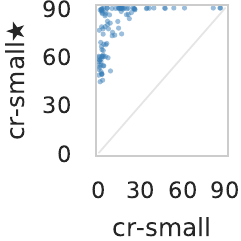}
	}
	\subfloat[\stress]{
		\includegraphics{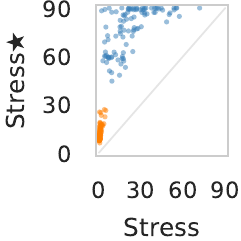}
	}
	\subfloat[\frcos]{
		\includegraphics{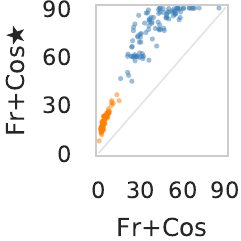}
	}
	\caption{Initial crossing angle vs the final crossing angle. The plots show
		the crossing angles of the classes \rome and \TriX.}
	\label{fig:ivf:direct_apx}
\end{figure}

\begin{figure}
	\centering
	\subfloat[\rnd vs \rndRS \label{fig:ivf:random}]{
		\includegraphics{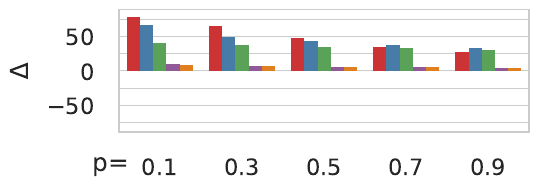}
	}
	\quad
	\subfloat[\frcos vs \frcosRS \label{fig:ivf:frcos}]{
		\includegraphics{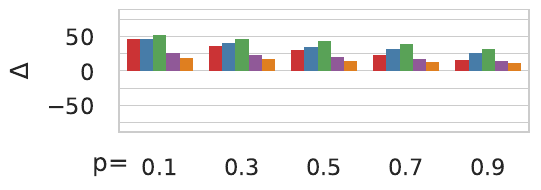}
	}

	\subfloat[\stress vs \stressRS \label{fig:ivf:stress}]{
		\includegraphics{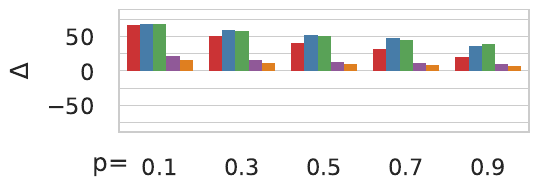}
	}
	\quad
	\subfloat[\crmin vs \crminRS\label{fig:ivf:crmin}]{
		\includegraphics{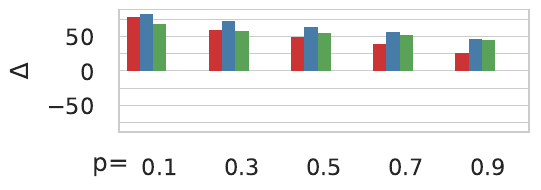}
	}

	\caption{Drawings of the initial sets vs drawings after the application of the
	random sampling approach.}
	\label{fig:ivf}

\end{figure}

\clearpage
\section{Increasing the Number of Iterations}
\label{sec:4n_Iterations}

In this section we compare the following two drawings. We start with \stress
drawings and apply $4n$ iterations with the \randomSampling approach with
the \Precise configuration (\stressRSS). Second, we apply $2n$ iterations to the \frcos
drawings. The plots in Fig.~\ref{fig:4n_iteration} show that drawings of graphs
in \north and \rome have indeed a slightly larger crossing angle when they are
obtained from \stress. For the remaining graph classes the drawings obtained
\frcos still have an advantage over \stress.
 
\begin{figure}[h]
	\centering
	\subfloat[]{
		\includegraphics{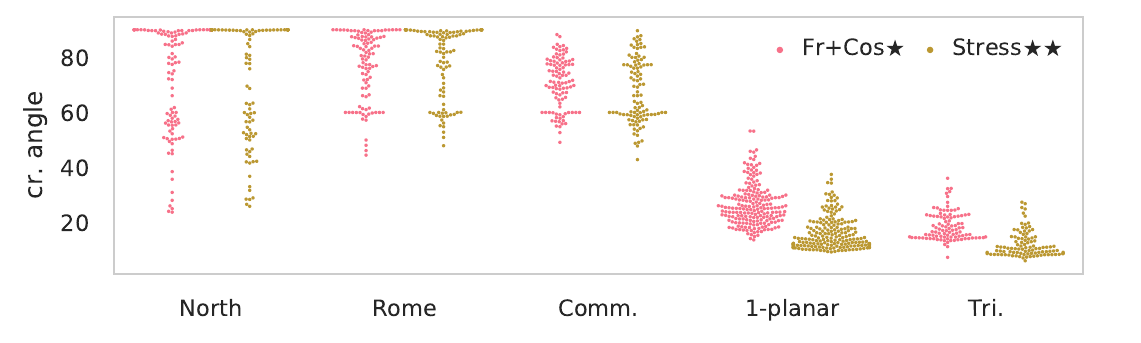}
	}
	
	\subfloat[]{
		\includegraphics{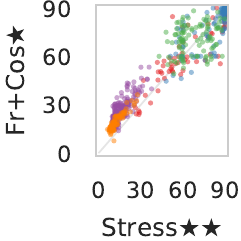}
	}
	\quad
	\subfloat[\frcosRS vs \stressRSS]{
		\includegraphics{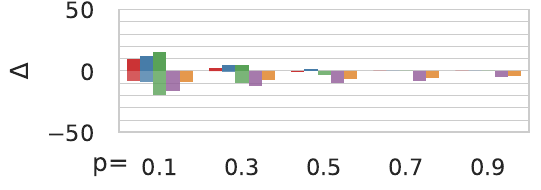}
	}
	\caption{Comparison}
	\label{fig:4n_iteration}
\end{figure}

\clearpage

\section{Number of Crossings vs Number of Operations}
\label{sec:cr_vs_ops}

In this section, we further evaluate our speed-up heuristic introduced
in Section~\ref{sec:fast_min_angle_compute}. We compare the number of crossings
$C$ in a drawing with the number $P$ of edge-pairs that the speed-up heuristics tests
for intersections. Ideally, $P$ is significantly smaller than $C$.

One data point in Fig.~\ref{fig:cr_vs_ops} corresponds to a single graph.  We
applied the $\Sloppy$ heuristic to the random drawings of all our benchmark
graphs. We counted in each iteration the number of crossings of the current
drawing and the number of tested pairs. The values in the figure 
correspond to the average of these values over $2n$ iterations. Note that the
plot uses a double log-scale.

We observe that for small instances, the heuristics tests more edge-pairs than
the drawing has crossings. With an increasing number of crossings, the heuristic
indeed tests less edge-pairs than crossings.

\begin{figure}[h]
	\centering
	\includegraphics{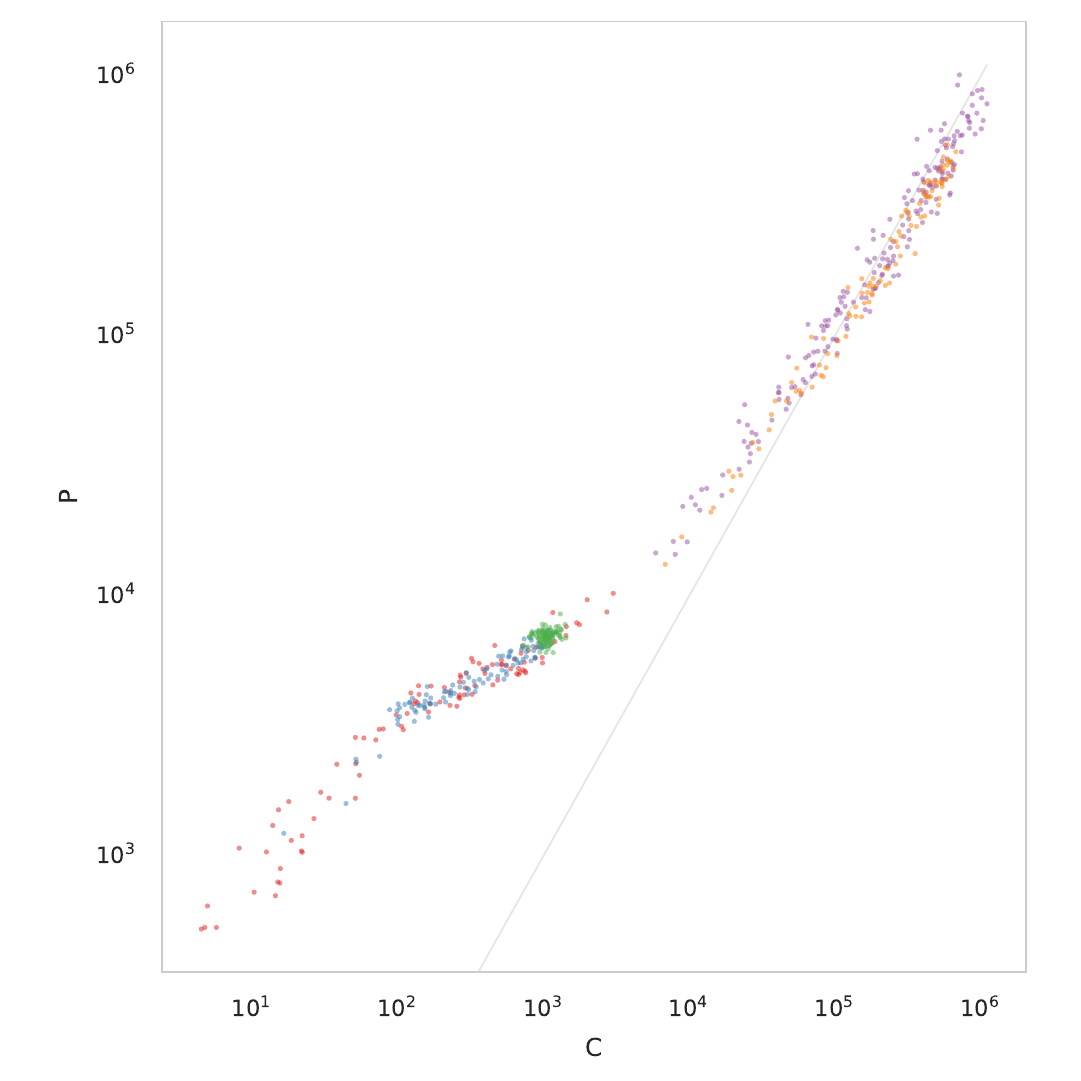}
	\caption{Number $C$ of crossings vs number $P$ of tested edge-pairs.}
	\label{fig:cr_vs_ops}
\end{figure}

\fi
\end{document}